\documentclass[aps,pra,amsmath,amssymb,onecolumn,superscriptaddress,11pt,notitlepage,showpacs]{revtex4-1} 
\usepackage{amsmath}
\usepackage{amsfonts}
\usepackage{amssymb}
\usepackage{graphicx}
\graphicspath{{./figures/}}
\usepackage{epsfig}
\usepackage{color}
\usepackage{epstopdf}
\newcommand{\tmpnote}[1]%
   {\begingroup{\color{blue} (FIXME: #1)}\endgroup}
\newcommand{\tmpnoteR}[1]%
   {\begingroup{\color{red} (FIXME: #1)}\endgroup}
 \newcommand{\comment}[1]%
       {\marginpar{\tiny C: #1}}

\begin{document}
\title{Cross-Kerr nonlinearity: a stability analysis}
\author{Roope Sarala}
\email[]{roope.s.sarala@student.jyu.fi}
\affiliation{Department of Physics and Nanoscience Center, University of Jyv\"askyl\"a,
P.O. Box 35 (YFL), FI-40014 University of Jyv\"askyl\"a, Finland}
\author{F. Massel}
\email[]{francesco.p.massel@jyu.fi}
\affiliation{Department of Physics and Nanoscience Center, University of Jyv\"askyl\"a,
P.O. Box 35 (YFL), FI-40014 University of Jyv\"askyl\"a, Finland}

\begin{abstract}
  We analyse the combined effect of the radiation-pressure and cross-Kerr
  nonlinearity on the stationary solution of the dynamics of a nanomechanical
  resonator interacting with an electromagnetic cavity. Within this setup, we
  show how the optical bistability picture induced by the radiation-pressure force
  is modified by the presence of the cross-Kerr interaction term. More
  specifically, we show how the optically bistable region, characterising the
  pure radiation-pressure case, is reduced by the presence of a cross-Kerr
  coupling term. At the same time, the upper unstable branch is extended by the
  presence of a moderate cross-Kerr term, while it is reduced for larger values
  of the cross-Kerr coupling. 
\end{abstract}

\maketitle

\hspace{0.5cm} {\footnotesize AMS numbers: 70k30,81v80 }

\section{Introduction}

In recent years, the exploration of the coupling between electromagnetic and
mechanical degrees of freedom has witnessed unprecedented interest both from the
applied and fundamental point of view (for a recent review see \cite{Aspelmeyer:2014ce}).
The physics of the optomechanical couplings has implications in fields as
diverse as the investigation of gravitational wave physics through, e.g. the
VIRGO \cite{Anonymous:kqVfZnE3} and LIGO \cite{Anonymous:j0Ww0aRI} consortia,
and the study of coherent quantum effects with a view to the manipulation of
quantum mechanical and optical degrees of freedom down to the level of single
quanta. Prominent examples of the achievements connected to the manipulation of
optomechanical degrees of freedom are represented by the ground-state cooling of
mechanical resonators \cite{Teufel:2011jga}, the optical/microwave transduction
by mechanical means \cite{Andrews:2014kja}, the demonstration of potential
quantum-limited amplification \cite{Massel:2011ca}.  

From this perspective, in cavity optomechanical systems strong emphasis has
lately been put to the so called single-photon strong coupling regime, for which
the dynamics of mechanical and optical degrees of freedom are strongly coupled,
leading e.g. to photon blockade effects \cite{Rabl:2011gn,Nunnenkamp:2011cp} or
the appearance of multiple sidebands in the optical spectrum of the cavity
\cite{Nunnenkamp:2011cp}.

Along these lines, it has recently been proposed \cite{Heikkila:2014hh} and
experimentally demonstrated \cite{Pirkkalainen:2015hh} that strong-coupling
between single-photons and mechanical quanta of motion can be achieved in a
microwave setup. In this setup, the addition of a strongly nonlinear inductive
element (single-Cooper-pair transistor SCPT) allows for an increase by several
orders of magnitude of the single-photon optomechanical coupling $g_0$. Beyond
the pure enhancement of the radiation-pressure (RP) coupling strength, in
\cite{Heikkila:2014hh} it has been shown how the inclusion of a nonlinear
circuit QED element determines the appearance of higher-order terms in the
effective interaction between optical and mechanical degrees of freedom. In this
context, it has therefore become relevant to discuss how higher-order
interaction terms modify the dynamical properties of optomechanical system (see
e.g. \cite{Seok:2013iw}), paralleling the analysis performed in different
setups, such as membrane in the middle geometries \cite{Jayich:2008iz} and,
analogously, in experiments exploiting optomechanical coupling with ultracold
gases \cite{Purdy:2010gh,Xuereb:2013ug}.

In this paper we discuss the role played by the lowest-order correction to the
interaction term beyond the RP term: the so-called cross-Kerr (CK) interaction
term. The CK term can be pictured as a change in the reflective index of the
cavity depending on the \textit{number} of phonons in the mechanical resonator,
while the RP term depends on the \textit{displacement} of the mechanical
resonator. In recent years, the cross-Kerr interaction has been shown to play a
crucial role in the field of quantum information processing; in particular in
the definition of near deterministic CNOT gates \cite{Nemoto:2004dr}, for
entanglement purification and concentration \cite{Sheng:2012jd,Sheng:2015dz}, in
the analysis of hyperentangled Bell states, and in the quantum teleportation of
multiple degrees of freedom of a single photon \cite{Sheng:2010ky}.

 We focus here on how the CK coupling affects the stability of the system in presence of a
strong driving coherent tone, focusing primarily on the optimally detuned
condition on the red sideband. Our main goal is to show how the CK term plays a
nontrivial role in the determination of the stability picture of the system
under consideration, generally extending the parameters window within which the
dynamics of the system is stable. 

\begin{figure}
\centering
\includegraphics[width=\textwidth]{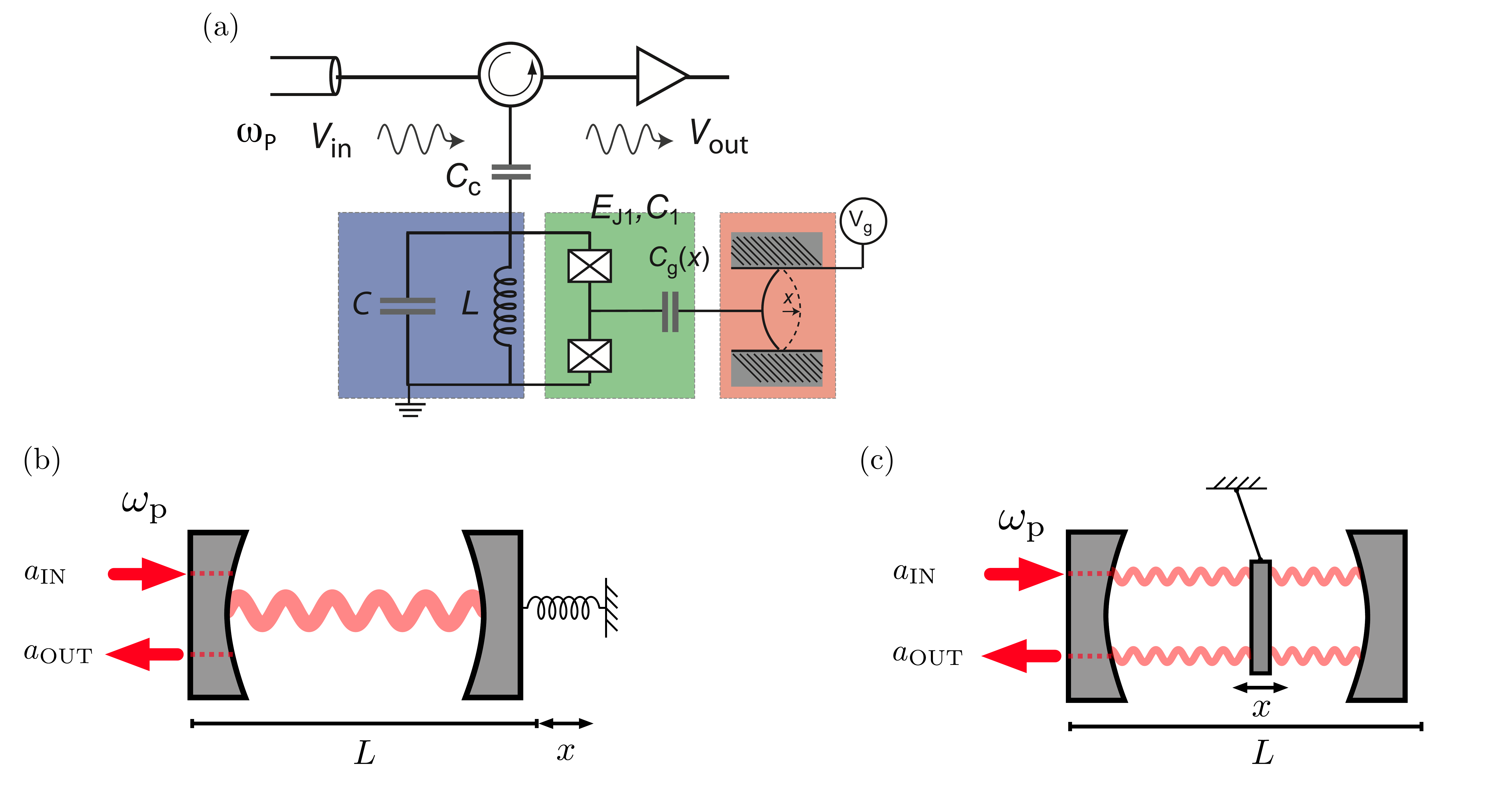}
\label{fig:thesetup}
\caption{Schematic picture of the system. (a) Lumped element representation of
  the system in presence of SCPT (from \cite{Pirkkalainen:2015hh}). (b) Regular
  optomechanical setup, and (c) membrane-in-the-middle setup.}
\label{fig:setups}
\end{figure}

\section{The model}

In our analysis, the dynamics of the system is generated by the following
Hamiltonian 
\begin{equation}
\label{eq:1}
H = H_0 + H_{\rm rp} + H_{\rm ck},
\end{equation}
where (hereafter $\hbar=1$)
\begin{equation}
  \label{eq:2}
  H_0 = \omega_{\rm c} a^{\dagger} a + \omega_{\rm m} b^{\dagger} b  
\end{equation}
is the Hamiltonian for the non-interacting electromagnetic cavity/mechanical
resonator system, where $\omega_{\rm c}$ and $\omega_{\rm m}$ denote the
resonance frequencies of the cavity and the mechanics, and $a$ ($a^{\dagger}$)
and $b$ ($b^{\dagger}$) are the boson annihilation (creation) operators for the
cavity and the mechanics respectively. As it is usually the case in the analysis
of optomechancial systems, we focus our attention on one cavity and one
mechanical resonator modes.  The RP coupling term $H_{\rm rp}$
is given by
\begin{equation}
  \label{eq:3}
  H_{\rm rp} = g_0 a^{\dagger} a \left( b^{\dagger} + b \right).
\end{equation}
The term $H_{\rm rp}$ can be understood as a shift of the cavity resonant
frequency as a function of the displacement of the mechanical resonator
$x \propto (b^\dagger+b)$. Depending on whether a Fabry-P\'erot cavity or a
lumped element circuit picture is adopted, according to the standard cavity
optomechanics description, the shift in the resonant frequency can be modeled
either as a change in the physical length of the cavity or as a change in the
capacitance induced by the displacement of the mechanical resonator (see
Fig. \ref{fig:setups}). From both perspectives, the coupling constant $g_0$ can
thus be expressed as 
\begin{equation}
  \label{eq:3.1}
  g_0=x_{\rm zpf}  \frac{\partial \omega_{\rm c}(x)}{\partial x} 
\end{equation}
where $x_{\rm zpf}$ is the zero-point potion associated with the displacement of
the mechancal resonator.
In the lumped element description, the resonant frequency  given by $\omega_{\rm
c}=1/\sqrt{L C(x)}$, where, as mentioned above, we have allowed for the possibility of a
mechanical-position dependence of the overall circuit capacitance, leading thus
to the following expression
\begin{equation}
  \label{eq:4}
  g_0=-x_{\rm zpf} \frac{\omega_{\rm c}}{2C} \frac{\partial C(x)}{\partial x} 
\end{equation}

In \cite{Pirkkalainen:2015hh}, it has been proposed and experimentally verified
that the introduction of a strongly nonlinear circuit element (the SCPT) will
boost the optomechanical coupling by several orders of magnitude. Essentially
the RP coupling between cavity and mechanics, in presence of the Josephson
junction qubit, is mediated by the strong nonlinear inductance determined by the
presence of the qubit.  Furthermore, in \cite{Heikkila:2014hh} it was shown that
the qubit-mediated coupling introduces higher-order terms in the coupling
between optics and mechanics.  In this paper we focus on the lowest-order
coupling term relevant in the pump/probe setup, ubiquitous in the analysis of
optomechanical systems: from \cite{Heikkila:2014hh}, invoking the rotating-wave
approximation, this term is given by
\begin{equation}
H_{\rm ck} = g_{\rm ck} a^{\dagger} a b^{\dagger} b,
\end{equation}
where 
\begin{equation}
  \label{eq:6}
  g_{\rm ck}=x_{\rm zpf}^2 \frac{\partial^2 \omega_{\rm c}}{\partial x^2}
\end{equation}
As previously mentioned, this term can be interpreted as an electromagnetic
field phase shift induced by the number of mechanical phonons, as opposed to the
RP term, which represents a phase shift of the field, depending
on the displacement of the mechanical resonator from the equilibrium position.

In our analysis, we describe the coupling between the optomechanical system and
the environment in terms of input-output formalism \cite{Walls:1105914}. Within
this framework, the equations of motion generated by the Hamiltonian
(\ref{eq:1}) for $a$ and $b$ can be written as

\begin{align}
\dot{a} &= - i\left[\omega_{\rm c} +
                                      g_{\rm 0}\left( b^{\dagger} + b \right) + 
                                      g_{ \rm ck}b^{\dagger}b 
                               \right] a 
                           - \frac{\kappa}{2}a 
                           + \sqrt{\kappa}a_{\rm in} \label{eq:eom1} \\
\dot{b} &= -i\left( \omega_{ \rm m} + 
                                      g_{\rm ck}a^{\dagger}a 
                               \right) b - g_{\rm  0}a^{\dagger}a
                          - \frac{\gamma}{2} b + \sqrt{\gamma}b_{\rm in}, \label{eq:eom2}
\end{align}
where $\kappa$ and $\gamma$ describe the coupling to the environment of the
cavity and the mechanical degrees of freedom, respectively.

Adopting a standard approach, with a view to considering a situation in which
the cavity is strongly driven by a coherent signal
$\alpha_{\rm in} \exp\left[-i \omega_{\rm p} t\right]$ and in absence of a direct
driving of the mechanical motion, we decompose the field operators $a$, $b$,
$a_{\rm in}$ and $b_{\rm in}$ so that $a = \alpha + \delta a$,
$b = \beta + \delta b$,
$a_{\rm in} = \alpha_{\rm in} + \delta a_{\rm in}$ and
$ b_{\rm in} = \beta_{\rm in} + \delta b_{\rm in}$. Assuming that the
amplitude of the fluctuations to be small with respect to the coherent signals,
in a frame co-rotating with the input coherent drive at frequency
$\omega_{\rm p}$, we can write the zeroth-order approximation to
Eqs. (\ref{eq:eom1}) and (\ref{eq:eom2}) as
\begin{align}
  \alpha &= \frac{ \sqrt{\kappa}\alpha_{\rm in} }{ \frac{\kappa}{2} -i \left[
           \Delta_0 -g_0 \left( \beta^* + \beta \right)- g_{\rm ck} |\beta|^2 \right] } \label{eq:sseom1} \\
  \beta &= - \frac{  g_0 |\alpha|^2 }{\frac{\gamma}{2}  + i \left( \omega_{\rm m} + g_{\rm ck}|\alpha|^2 \right) }, \label{eq:sseom2}
\end{align}
where we have introduced the detuning
$\Delta_0 = \omega_{\rm p} - \omega_{\rm c}$ and we have assumed
$\beta_{\rm in} = 0 $. The solutions of Eqs. (\ref{eq:sseom1}) and
(\ref{eq:sseom2}), as a function of the input field amplitude $\alpha_{\rm in}$,
represent the equilibrium solutions for the cavity field amplitude $\alpha$ and
for the amplitude of the mechanical oscillator $\beta$, in a frame rotating at
$\omega_{\rm p}$. For $g_{\rm ck}=0$, Eqs. (\ref{eq:sseom1}) and
(\ref{eq:sseom2}) reproduce the steady-state solution obtained for a regular
optomechanical system in presence of RP coupling
\cite{Aldana:2013iya, Aspelmeyer:2014ce}. In that case, if the input field is
detuned so that $|\Delta| = \omega_{\rm m}$ allowing for the optimal exchange of
energy between cavity and the mechanical modes, in particular
$\Delta_0 = -\omega_{\rm m}$ represents the optimal cooling condition (see
\cite{Marquardt:2007dn, Teufel:2011jga}), while $\Delta_0 = \omega_{\rm m}$
allows for optimal amplification of an incoming signal around the cavity
resonant frequency \cite{Massel:2011ca}. It has recently been shown
\cite{Khan:2015kf} that, in presence of CK coupling and for moderate driving,
such as to allow for the use of pure RP steady-state solutions,
the optical damping of the mechanics is a non-monotonous function of the optical
drive, thus potentially hindering the cooling of the mechanical motion by
optical means for a red-detuned pump, and on the other, limiting the parametric
instability for a blue sideband drive.

In the present work we discuss a different aspect of the problem, by considering
the stability properties of the solutions of Eqs. (\ref{eq:sseom1}) and
(\ref{eq:sseom2}). The determination of the stability of the system is performed
through the standard stability analysis of linear autonomous systems of
differential equations, allowing us to identify the stability character of the
equilibrium solutions of Eqs. (\ref{eq:sseom1}) and (\ref{eq:sseom2}) as a
function of the different parameters characterising the system, with particular
focus on $g_{\rm ck}$.




\section{Steady-state solutions}
\label{sec:ss-sol}

From Eqs. (\ref{eq:sseom1}) and (\ref{eq:sseom2}), it is clear how the CK
interaction can modify the optical bistability picture associated with the RP
only setup (Fig.  \ref{fig:Fig2}): due to the presence of higher-order terms
both in $\alpha$ and $\beta$, in principle, a more complex stability diagram as
a function of the external drive should be expected, more specifically a larger
number of (stable or unstable) equilibrium points should appear. However, while
some important qualitative and quantitative differences do arise because of the
presence of the CK term, no extra physical equlibrium solutions are actually
present. More specifically, combining Eqs. (\ref{eq:sseom1}) and
(\ref{eq:sseom2}), we obtain a quintic equation for $|\alpha|^2$, which
represents the mean-field cavity occupation -- we note here that, for pure RP
($g_{\rm ck} = 0$), the equation would be cubic. Moreover, both for the pure RP
case and in presence of CK coupling, $|\alpha|^2$ can have either one or three
positive and real solutions, meaning that in the quintic equation associated
with the CK case, two of the solutions violate the condition $|\alpha|^2\geq 0$
for all values of the parameters, and thus have to be discarded.

For positive detunings $\Delta_0 > 0$, all values of $|\alpha|^2$ corresponding
to the solutions of Eqs.  (\ref{eq:sseom1}) and (\ref{eq:sseom2}) would be
complex or negative, therefore implying that the optical bistability can only be
found for negative detunings (red sideband, see also \cite{Aldana:2013iya}).
The values of $\alpha_{B}$ and $\alpha_{C}$, along with the corresponding
input-field values $\alpha_{\rm in}^{B}$ and $ \alpha_{\rm in}^{C}$ points
define the branch $\mathbf{B}-\mathbf{C}$ in Fig. \ref{fig:Fig2} (a).  As
previously mentioned, in this case, the presence of the nonzero CK term, due to
considerations concerning the unphysical nature of some of the solutions to eqs.
(\ref{eq:sseom1}) and (\ref{eq:sseom2}), does not introduce further equilibrium
points in the dynamical system analysis.  However, as it is possible to see from
Fig. \ref{fig:Fig2}, the presence of a CK coupling term shifts the relative
position of points ${\rm B}$ and ${\rm C}$, reducing the range of pump amplitude
$\alpha_{\rm IN}$ for which the branch $\mathbf{B}-\mathbf{C}$ is present. 
In order to discuss the stability character of the steady-state solutions, we
investigate the stability properties of the following dynamical system
\begin{align}
    \delta \dot{a} &= i \Delta \delta a  -\frac{\kappa}{2} \delta a + i G
  (\delta b + \delta b^\dagger) + \sqrt{\kappa} \delta a_{\rm in} \nonumber \\
   \delta \dot{b} &= -i \omega^e_{\rm m} \delta b  -\frac{\gamma}{2} \delta b + i (G^*
  \delta a + G \delta a^\dagger) + \sqrt{\gamma}\delta b_{\rm in}
 \label{eq:12}
\end{align}
which represent the first-order expansion of Eqs. \eqref{eq:eom1} and
\eqref{eq:eom2} around the equilibrium points defined by Eqs. \eqref{eq:sseom1}
and \eqref{eq:sseom2}, and are formally analogous to the first-order equations
obtained for the pure RP case (see e.g. \cite{Aspelmeyer:2014ce}). The essential
difference between the RP and the CK case, is that the linearised coupling $G$,
the effective detuning $\Delta$ and the effective mechanical frequency
$\omega^e_{\rm m}$ are given by the following expressions
\begin{align}
    G&=g_0 \alpha \left(1+ \frac{g_{\rm ck} |\alpha|^2}{\omega_{\rm m}+g_{\rm ck}
  |\alpha|^2}\right) \nonumber \\
  \Delta& = \Delta_0 + 2 \frac{g_0^2}{\omega_{\rm m}} |\alpha|^2 \left[1 + 
                                         \frac{g_{\rm ck} |\alpha|^2}
                                                 {\left(\omega_{\rm m}+g_{\rm ck}
  |\alpha|^2\right)^2}\right] \nonumber \\
  \omega^e_{\rm m}&=\omega_{\rm m}+g_{\rm ck}\left|\alpha\right|^2
\label{eq:14} 
\end{align}

\section{Stability analysis - lower branch}
\label{sec:stab-an}

In this section we study the stability character of the solutions of the cavity
optomechanical system, within the usual framework of stability analysis for
autonomous linear systems. More explicitly, the nonlinear differential equations
system given by (\ref{eq:eom1}) and (\ref{eq:eom2}) can be solved perturbatively
order-by-order. The solution of the zeroth-order equation corresponds to the
steady state solution given by Eqs. (\ref{eq:sseom1}) and (\ref{eq:sseom2}),
while the first-order equation of the fluctuations around the stationary
solutions, given by Eqs. \eqref{eq:12}, can be written as 
\begin{align}
 \label{eq:linsyst}
  \delta \dot{\mathbf{v}} = A \mathbf{v} +\delta \mathbf{v}_{\rm IN}
\end{align}
where $\delta  \mathbf{v}=\left[\delta a, \delta a^\dagger,\delta b, \delta
  b^\dagger \right]^T$, $\delta  \mathbf{v}_{\rm IN}=\left[\delta a _{\rm IN}, \delta a _{\rm IN}^\dagger,\delta b _{\rm IN}, \delta
  b _{\rm IN}^\dagger \right]^T$, and A is given by 
\begin{align}
A=\begin{bmatrix}
   i \Delta -\frac{\kappa }{2}& 0 & i G & i G \\
  0 & -i \Delta -\frac{\kappa }{2}  & -i G^* & -i G^* \\
  i G   &  i G^* & -i \omega_{\rm m} -\frac{\gamma }{2} & 0  \\
 -i G^* & -i G   & 0 & i \omega_{\rm m}-\frac{\gamma }{2}  \\
\end{bmatrix}
\label{eq:17}
 \end{align}

The stability character of the steady-state solution is then provided by the
sign of the solutions of the characteristic equation associated with $A$,
$p_A(\lambda)=0$.
It is possible to show how, from the Routh-Hurwitz criterion, the stability of  
the system is
characterised by the two following conditions,
\begin{align}
  \label{eq:8}
 &\frac{\gamma^2 \Delta^2}{4} +\frac{\gamma^2 \kappa^2}{16} + 4 g^2 \Delta
  \omega^e_{\rm m} +\Delta^2 {\omega^e_{\rm m}}^2 +\frac{\kappa^2 {\omega^e_{\rm m}}^2}{4} >0 \\
 & \left[(\gamma+\kappa) \Gamma_1 -\Gamma_2\right] \Gamma_2-(\gamma+\kappa)^2 \Gamma_3>0
 \label{eq:8.1}
\end{align}
with
\begin{align*}
\label{eq:9}
  & \Gamma_1= \frac{\gamma^2}{4}+\Delta^2 + \gamma \kappa + \frac{k^2}{4}+{\omega^e_{\rm m}}^2\\
  & \Gamma_2= \frac{\gamma ^2 \kappa }{4}+\gamma  \Delta ^2+\frac{\gamma  \kappa
    ^2}{4}+\kappa  {\omega^e_{\rm m}}^2\\
  & \Gamma_3=\frac{\gamma ^2 \Delta ^2}{4}+\frac{\gamma ^2 \kappa ^2}{16}+\Delta ^2
  {\omega^e_{\rm m}}^2+4 \Delta  |G|^2 \omega^e_{\rm m}+\frac{\kappa ^2 {\omega^e_{\rm
      m}}^2}{4}
\end{align*}
which correspond to the conditions $a_0>0$ and $a_3a_2a_1-(a_1^2+a_3^2a_0)>0$,
for the characteristic polynomial $p_A(\lambda)=\lambda^4 + a_3 \lambda^3+
a_2 \lambda^2+  a_1 \lambda+ a_0$, the other conditions for the stability of the
system being identically satisfied for the system under consideration.
\begin{figure}
\centering
\includegraphics[width=0.6\textwidth]{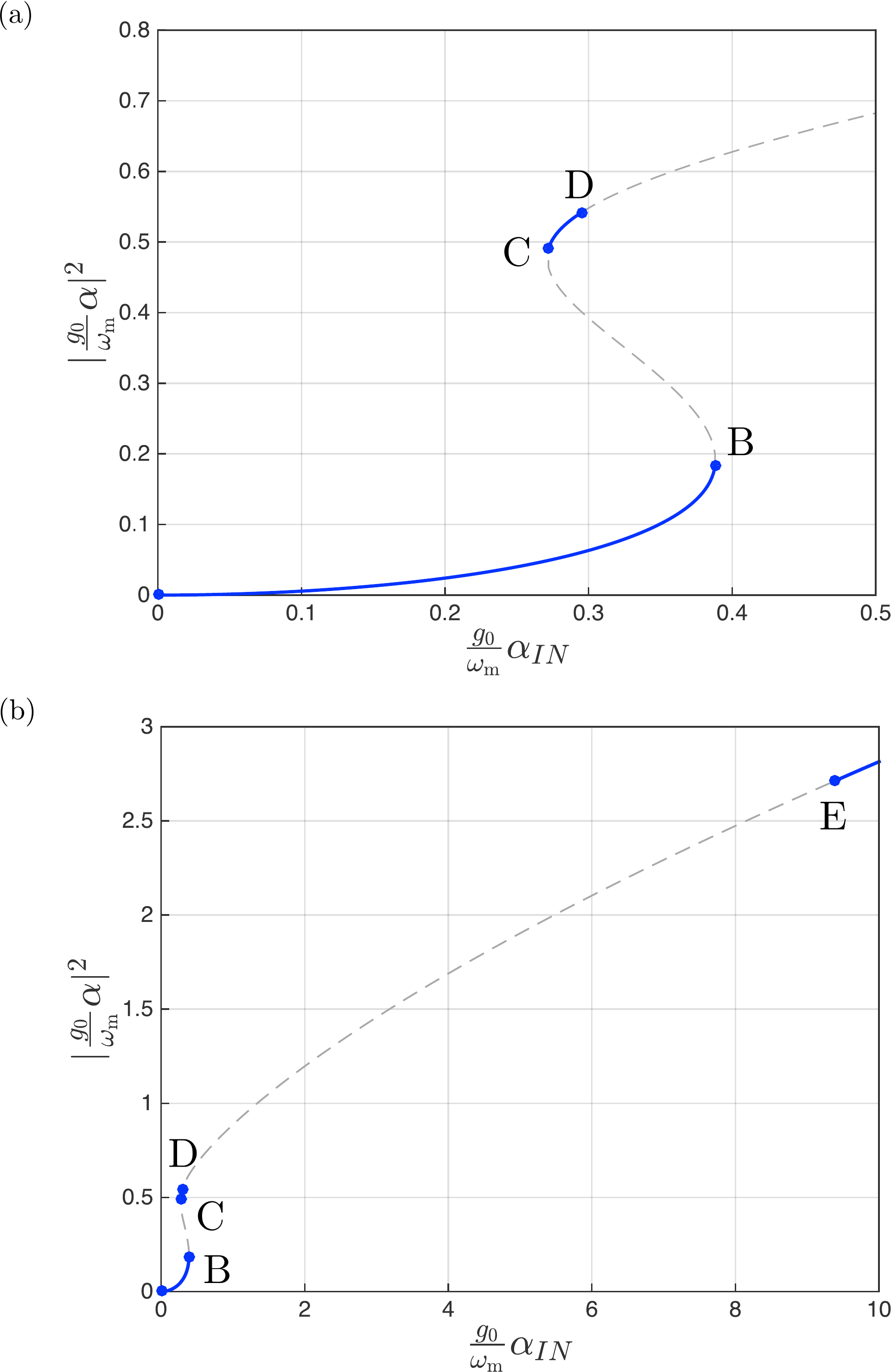}
\caption{Optical bistability in the RP only case. The unstable 
  branch $\mathbf{B}-\mathbf{C}$ corresponds to the lower branch, associated with the optical
  bistability picture (condition \eqref{eq:8}). The unstable branch $\mathbf{D}-\mathbf{E}$
  corresponds to the violation of condition \eqref{eq:8.1}. Plot parameters: 
  $\Delta_0=-\omega_{\rm m}$, $g_0=10^{-5}\omega_{\rm m} $, $\kappa=0.6
  \omega_{\rm m}$, $\gamma=0.12 \omega_{\rm m}$}
\label{fig:Fig2}
\end{figure}
The violation of the condition $a_0>0$ corresponds to the branch
$\mathbf{B}-\mathbf{C}$ . As outlined in \cite{Aldana:2013iya}, in the pure
RP case, up to linear order in $\gamma/\omega_{\rm m}$, the
critical values for $|\alpha|^2$, corresponding to the points $B$ and $C$ (see
Fig. \ref{fig:Fig2}) are given by
\begin{equation}
  \label{eq:7}
  \left|\alpha_{B,C}\right|^2=-\frac{\Delta_0}{3 \omega_{\rm m}}\left[1\pm \frac{1}{2}\sqrt{1-\frac{3
        \kappa^2}{4 \Delta_0^2}}\right],
\end{equation}
indexes $B$ and $C$ correspond to the lower and upper sign respectively.
In presence of CK coupling, the stability picture is modified. For small
values of $g_{\rm ck}$, the presence of a CK coupling increases the
range of values of $|\alpha|^2$ for which the branch $\mathbf{B}-\mathbf{C}$ is unstable. In the
resolved-sideband regime  ($\kappa \ll \omega_{\rm m}=|\Delta_0|$) the CK result
is obtained as a (first-order) expansion in $g_{\rm ck}$ from the pure RP
solution  
\begin{align}
  \label{eq:10}
  &|\alpha^{\rm ck}_{B}|^2=-\frac{\Delta_0}{6\omega_{\rm m}}-\frac{\kappa^2}{16\Delta_0 \omega_{\rm m}}-\frac{g_{\rm
  ck} \Delta_0}{4 \omega_{\rm m}^4}\left(\Delta_0^2+\frac{3\kappa^2}{4}\right) \\
   &|\alpha^{\rm ck}_{C}|^2=-\frac{\Delta_0}{2\omega_{\rm m}}+\frac{\kappa^2}{16\Delta_0 \omega_{\rm m}}-\frac{g_{\rm
  ck} \Delta_0}{4 \omega_{\rm m}^4}\left(\Delta_0^2-\frac{\kappa^2}{4}\right). 
\end{align}
However, increasing $g_{\rm ck}$ further, leads to a reduction and, eventually,
the disappearance of the unstable branch $\mathbf{B}-\mathbf{C}$ (see
Figs. \ref{fig:Fig3} and \ref{fig:Fig3b}).
\begin{figure}[H]
\centering
\includegraphics[width=0.6\textwidth]{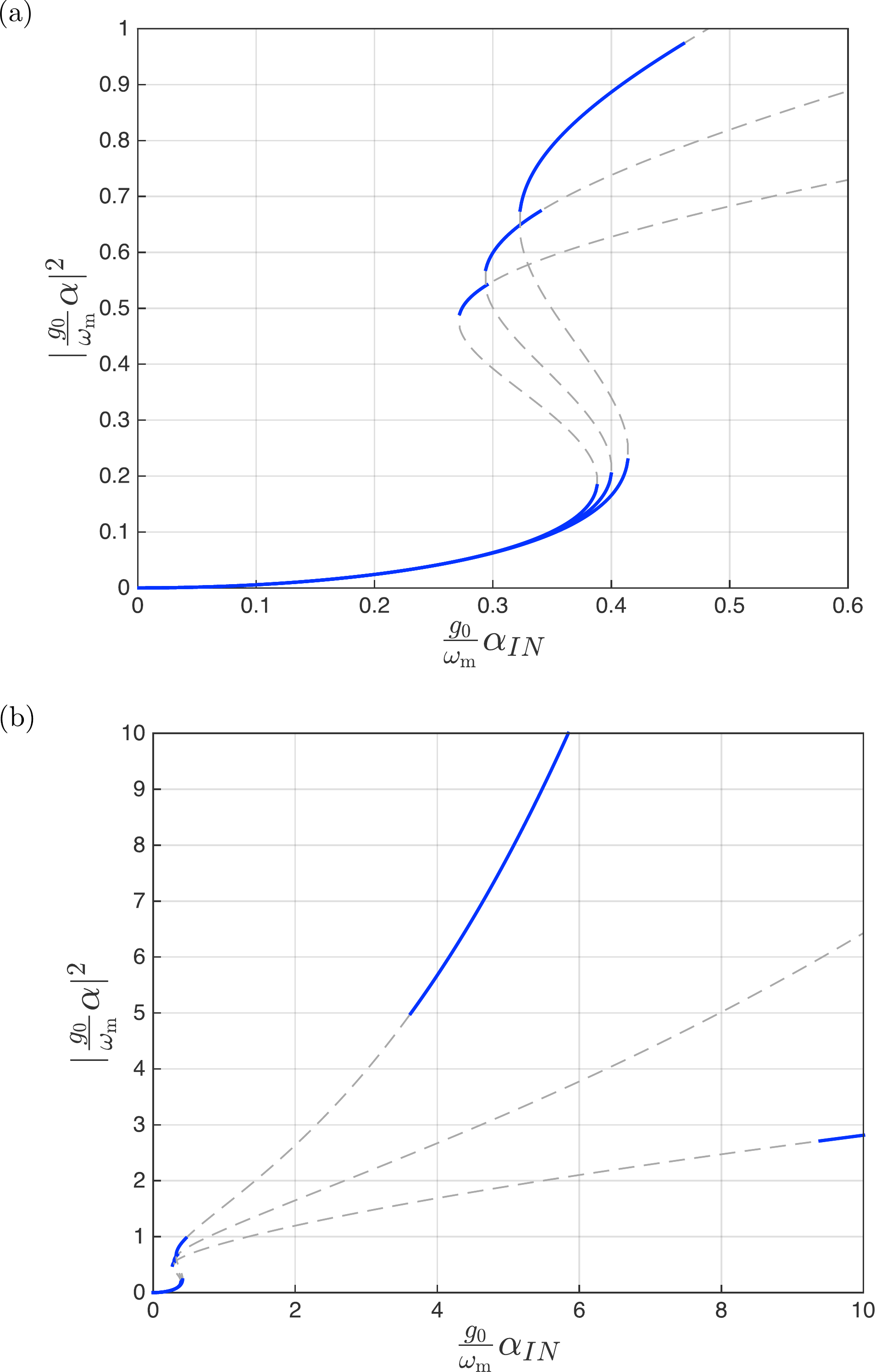}
\caption{Effect of $g_{\rm ck}$ on the stability diagram as a function of the
  input drive, weak $g_{\rm ck}$. (a) The presence of the CK term reduces the input drive range for
  the branch $\mathbf{B}-\mathbf{C}$ (increasing $g_{\rm ck}$ for the different curves from
  left to right). (b) The CK term first induces a broadening of the upper
  unstable region, which, upon further increase, is reduced (increasing
  $g_{\rm ck}$ for the different curves from bottom to top). Plots parameters,
  $g_{\rm ck}/g_0^2=0,0.2,0.4$; other parameters same as in Fig. \ref{fig:Fig2} }
\label{fig:Fig3}
\end{figure}
\begin{figure}[H]
\centering
\includegraphics[width=0.6\textwidth]{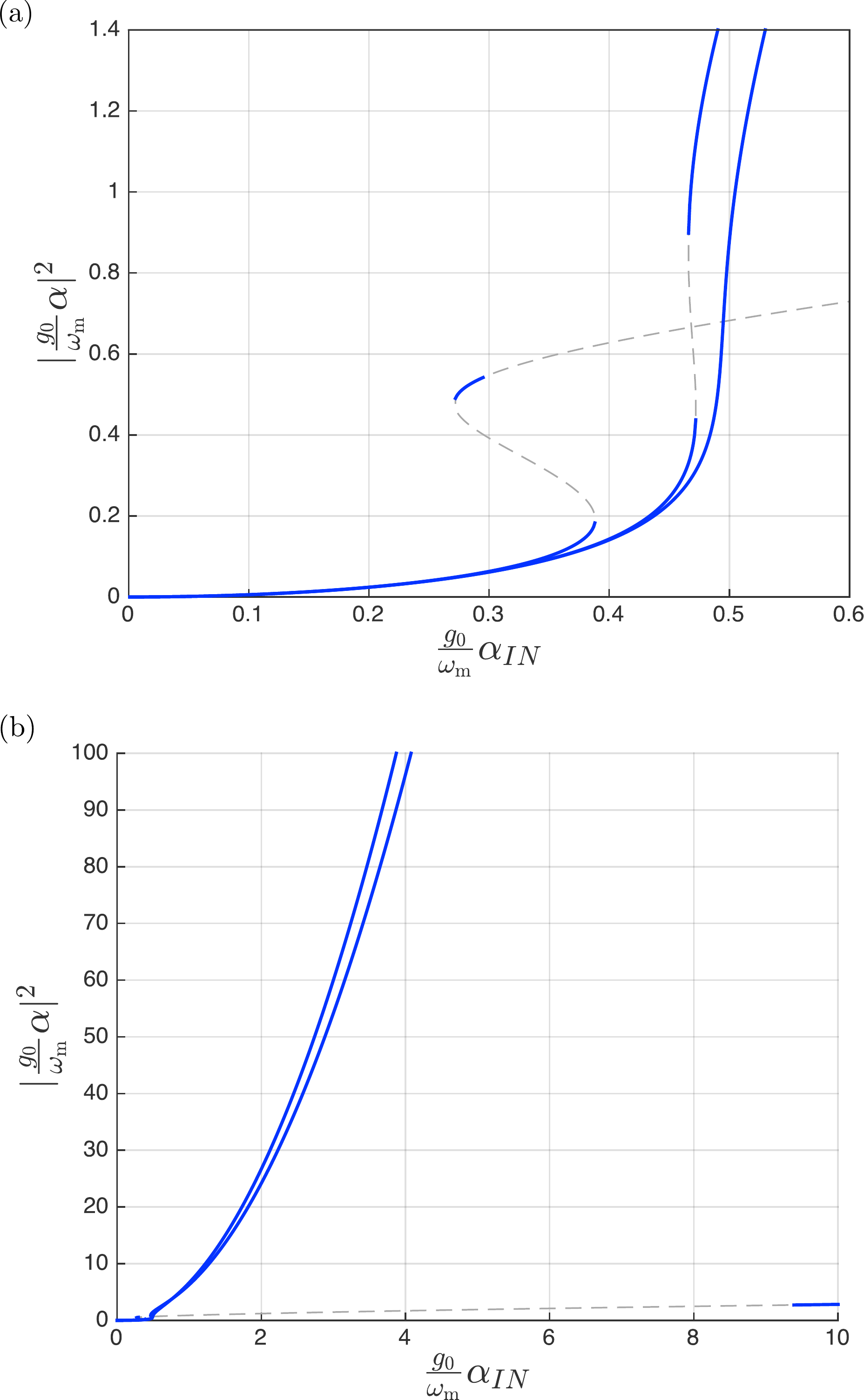}
\caption{Effect of $g_{\rm ck}$ on the stability diagram as a function of the
  input drive, strong $g_{\rm ck}$. (a) The presence of the CK term reduces the
  input drive range for the branch $\mathbf{B}-\mathbf{C}$ (increasing $g_{\rm ck}$ for the
  different curves from left to right). (b) The CK term first induces a
  broadening of the upper unstable region, which, upon further increase, is
  reduced (increasing $g_{\rm ck}$ for the different curves from bottom to
  top). Plots parameters, $g_{\rm ck}/g_0^2=0,0.95,1$ other parameters same as in
  Fig. \ref{fig:Fig2} }
\label{fig:Fig3b}
\end{figure}
In the regime for which $g_{\rm ck} |\alpha|^2\gg 1$, the critical value
$g^{\rm c}_{\rm ck}$ for which the unstable branch $\mathbf{B}-\mathbf{C}$ disappears can be
approximated by
\begin{align}
  \label{eq:11}
  g^{\rm c1}_{\rm ck}=\frac{8 \omega_{\rm m}^2}{\kappa} \sqrt{16-\frac{\gamma^2}{4\omega_{\rm m}^2}}
\end{align}
 
\section{Stability analysis - upper branch}
\label{sec:stab-analys-upper}

The condition given by Eq. \eqref{eq:8}, derived from the condition
$a_3a_2a_1-(a_1^2+a_3^2a_0)>0$, corresponds to the unstable branch
$\mathbf{D}-\mathbf{E}$. For the pure RP case, the endpoints of the
$\mathbf{D}-\mathbf{E}$ branch are (up to linear order in $\gamma/\kappa$ and
$\Delta_0=-\omega_{\rm m}$)
\begin{align}
  \label{eq:5}
  & \left|\alpha_D\right|^2=g_0^{-2}\left(\frac{1}{2} +\frac{\epsilon}{8}\right) \\
  & \left|\alpha_E\right|^2 =g_0^{-2}\left(\sqrt{\frac{1}{2 \epsilon}}+\frac{3}{4} +\frac {19}{32} \sqrt{2\epsilon} - \frac{ \epsilon}{16}\right).
\end{align}
(where $\epsilon=\gamma/\kappa$)
Analogously to the approach employed to evaluate the effect of the CK term on
the $\mathbf{B}-\mathbf{C}$ branch in the small $g_{\rm ck}$ limit, we consider the perturbative
correction to the RP result induced by $g_{\rm ck}$. Up to linear order, the
boundaries of the unstable region $\mathbf{D}-\mathbf{E}$, for small values of $g_{\rm ck}$ (
again for $\gamma/k \ll 1$) are given by
\begin{align}
  \label{eq:13}
 & \left|\alpha^{\rm ck}_{\rm D}\right|^2=\frac{(1+\eta_{\rm D})^2}{2 g_0^2(1+3\eta_{\rm D}+\eta_{\rm D}^2)} \\
  & \left|\alpha^{\rm ck}_{\rm E}\right|^2=\frac{3(1+\eta_{\rm E})^2}{32  g_0^2 \Lambda_{\rm E}(1+2\eta_{\rm E})^2}
                 \left(\sqrt{\frac{3\kappa}{\gamma\Lambda_{\rm E}}} +3 \right) 
\end{align}
with 
\begin{align*}
    & \Lambda_{\rm D,E}=\frac{3(1+\eta_{\rm D,E})(1+2\eta_{\rm D,E})-\eta_{\rm D,E}^2}
                                                        {8(1+\eta_{\rm D,E})(1+2\eta_{\rm D,E})^2} \\
   &\eta_{\rm D,E}=\frac{g_{\rm ck}}{g_0^2}|\alpha_{\rm D,E}|^2. 
\end{align*}

Within the validity limit of the perturbative approximation, the unstable region
$\mathbf{D}-\mathbf{E}$ is extended by the presence of the CK nonlinearity.  Our numerical
analysis shows that the region $\mathbf{D}-\mathbf{E}$ increases with increasing $g_{\rm ck}$
until a critical value $g_{\rm ck}^*$, for which $\mathbf{E} \to \infty$, implying that,
for $g_{\rm ck} =g_{\rm ck}^*$, the upper branch is unstable for every
$\left|\alpha\right|>\left|\alpha_{\rm D}\right|$. The value of $g_{\rm ck}^*$
can be evaluated considering, in the limit $|\alpha|^2 \gg
1$, the condition given by Eq. \eqref{eq:8.1} is expressed by a fourth-order
polynomial in $\left|\alpha\right|^2$, $Q(\left|\alpha\right|^2)$. Furthermore,
the asymptotic behaviour of $Q(\left|\alpha\right|^2)$ can be deduced from the
sign of the highest-power coefficients $c_3^\infty$ and $c_4^\infty$.
In this limit, we have
\begin{align}
  \label{eq:15}
 & c^\infty_3= -\frac{32}{\omega_{\rm m}} \left[  \frac{\gamma \kappa \Delta_0}{\omega_{\rm m}^2} +\frac{g_{\rm ck}}{g_0^2}
  \left\{\frac{g_{\rm ck}\gamma \kappa\Delta_0}{4 g_0^2}+
   \left(\gamma+\kappa \right)  \right\} \right]   \\
 & c^\infty_4=\gamma \kappa \left( \frac{g_{\rm ck}^2}{g_0^4}
   -\frac{4}{\omega_{\rm m}^2}  \right)^2
\end{align}
If we assume $2 \Delta_0 >g_{\rm ck}$, it is possible to see how
$c^\infty_3<0$ for every value of the cavity field, while $c^\infty_4>0$, except
when $g_{\rm ck}=2 g_0^2/\omega_{\rm m}$. In this case $\left|\alpha_{\rm E}\right|^2 \to
\infty$, and the system remains unstable for every value of the cavity field above 
$\left|\alpha_{\rm D}\right|$. If $g_{\rm ck}$ is further increased above
$g_{\rm ck}^*=2 g_0/\omega_{\rm m}$ the region $\mathbf{D}-\mathbf{E}$ becomes finite again and eventually
disappears for $g_{\rm ck}^{\rm c2}$. In the limit $\gamma/\kappa \ll 1$, 
\begin{align}
  \label{eq:16}
  g_{\rm ck}^{\rm c2}\simeq -\frac{2(\gamma+\kappa)^2}{\gamma \kappa
  \Delta_0}g_0^2 \simeq
  \left. 2 g_{\rm ck}^* + \frac{2 \kappa g_0^2}{\omega_{\rm m}\gamma} \right|_{\Delta_0 =-\omega_{\rm m}}.
\end{align}

\begin{figure}[H]
\centering
\includegraphics[width=0.6\textwidth]{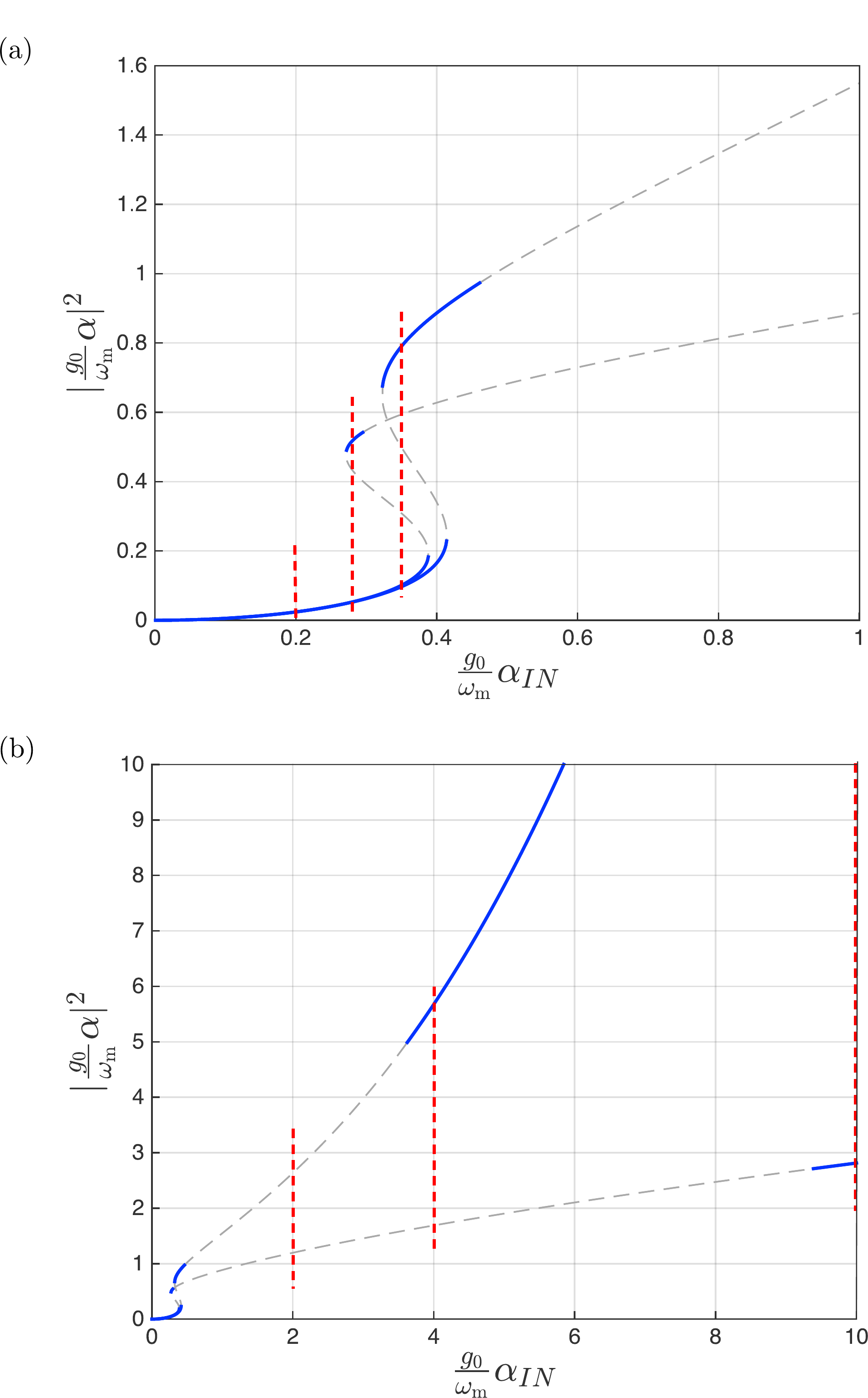}
\caption{Same as in Fig. \ref{fig:Fig3}, for the puer RP case ($g_{\rm ck}=0$),
  and for $g_{\rm ck}=0.2$. The vertical dashed lines correspond to the input
  drive values for which the $\Delta_0$-dependence is depicted in
  Figs. \ref{fig:Fig5}-\ref{fig:Fig7}.}
\label{fig:Fig4}
\end{figure}

\begin{figure}[H]
\centering
\includegraphics[width=0.6\textwidth]{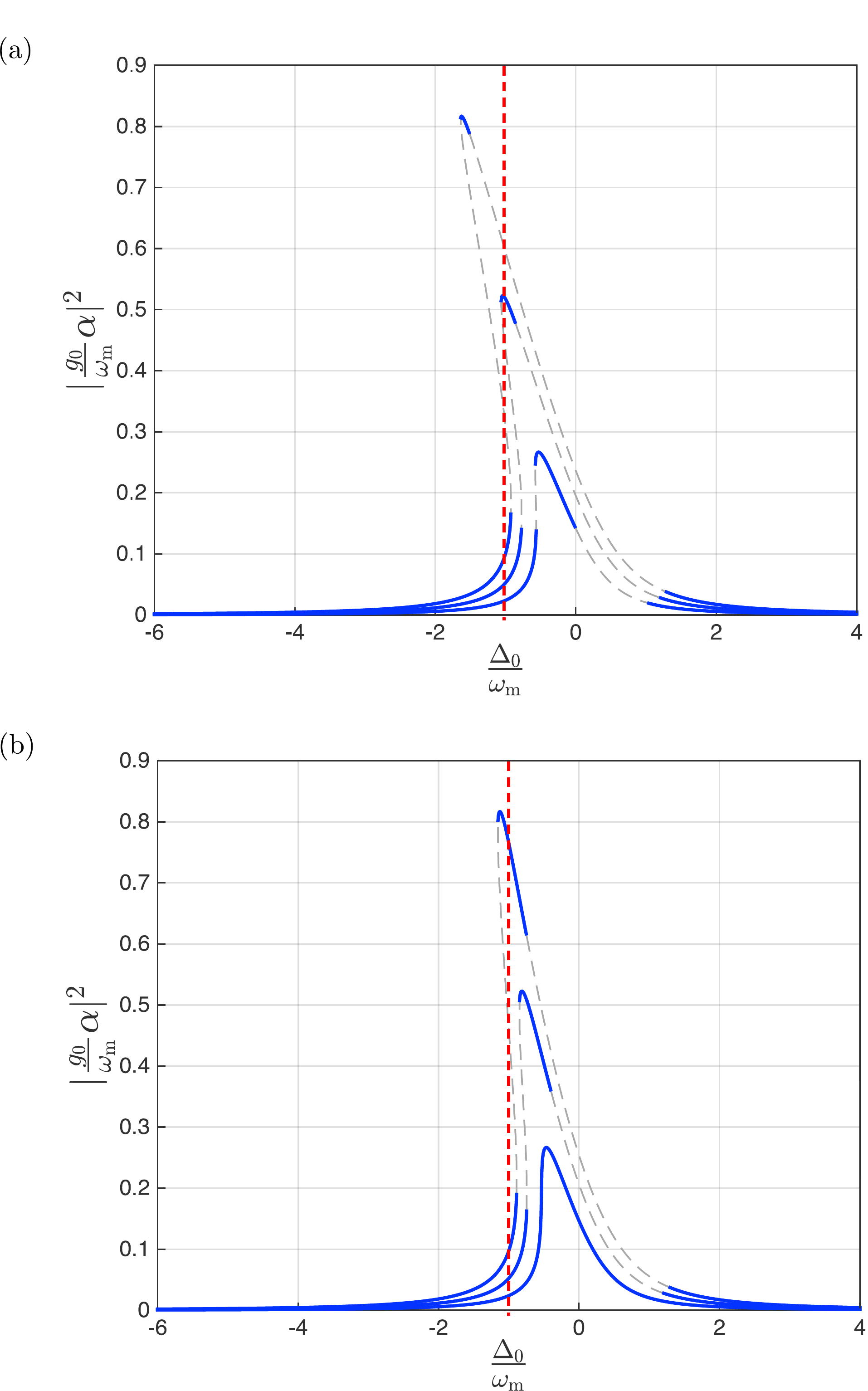}
\caption{$\Delta_0$ dependence of the cavity population 
in (a) the pure RP  case and (b) in the CK ($g_{\rm ck}=0.2$) case. 
Increasing $\alpha_{\rm IN}= 0.2, 0.28, 0.35$ in units of $\omega_{\rm m}/g_0$ for increasing
peak height. The presence of the CK term leads to a larger parameter range for
which stability is observed.}
\label{fig:Fig5}
\end{figure}

\begin{figure}[H]
\centering
\includegraphics[width=0.6\textwidth]{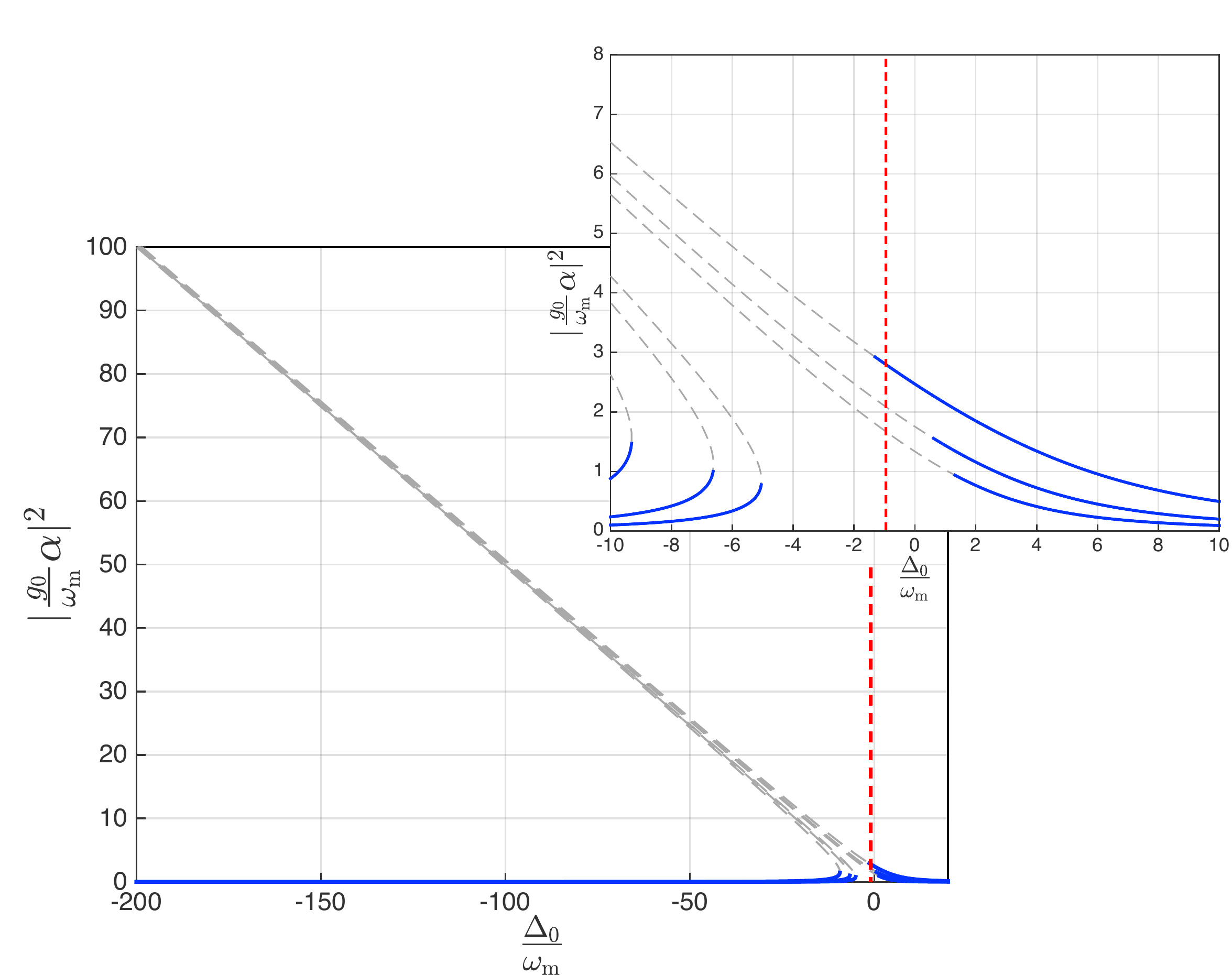}
\caption{$\Delta_0$ dependence of the cavity population in the pure RP
  case. Increasing $\alpha_{\rm IN}=2,4,10$ in units of $\omega_{\rm}/g_0$ for
  increasing peak height (see inset). The case $\alpha_{\rm IN}=10 \omega_{\rm}/g_0$ outlines the
  mechanism thorough which the result for $\Delta_0=-\omega_{\rm}$ is stable at
  large drives (vertical dashed line).}
\label{fig:Fig6}
\end{figure}

\begin{figure}[H]
\centering
\includegraphics[width=0.6\textwidth]{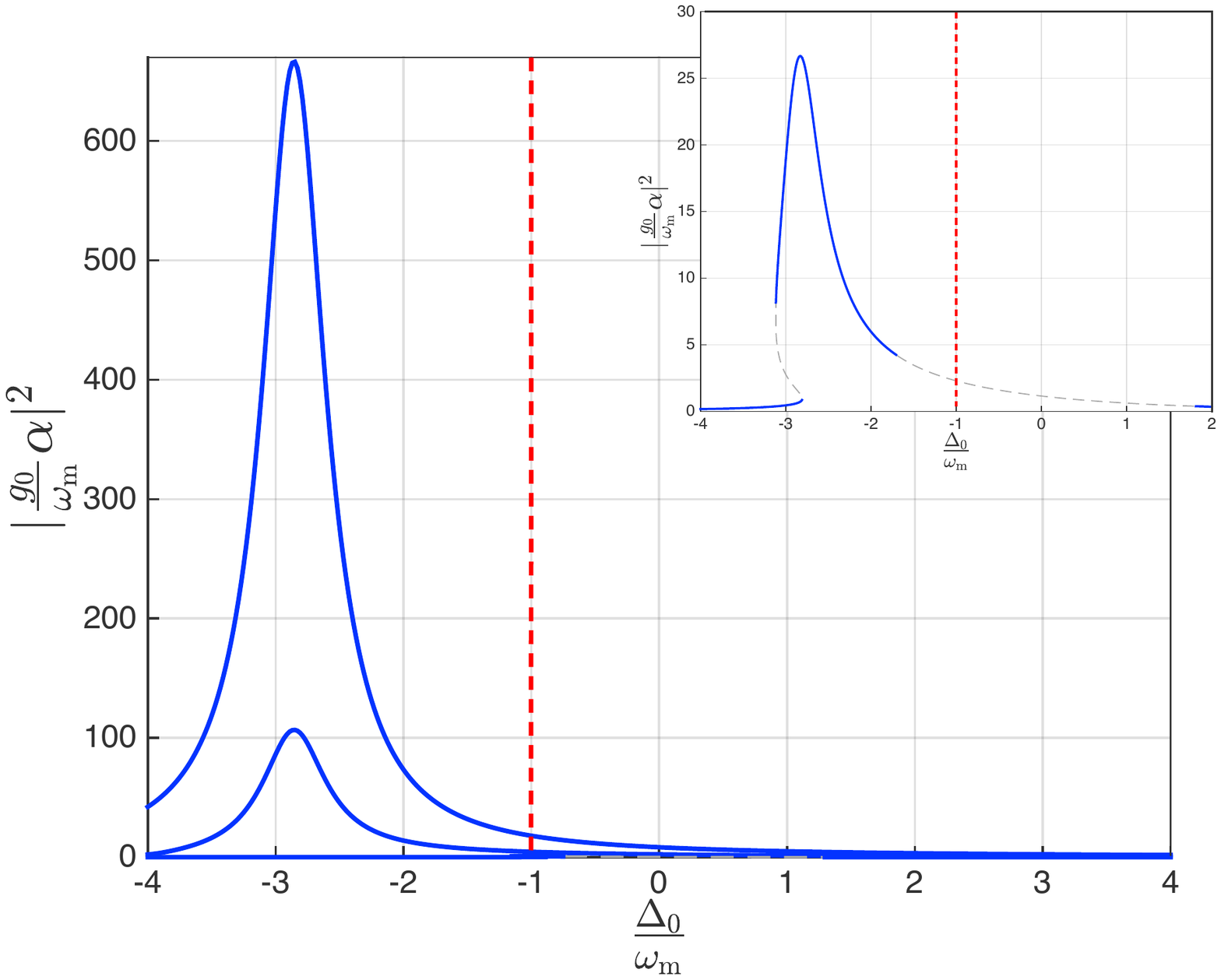}
\caption{$\Delta_0$ dependence of the cavity population in the CK
  ($g_{\rm ck}=0.2$) case.  Increasing $\alpha_{\rm IN}=$
  ($2,4,10 \omega_{\rm}/g_0$) for increasing peak height. It is possible to see
  how, for large values of the input field, the system is stable for every value
  of detuning $\Delta_0$.  The case $\alpha_{\rm IN}=2 \omega_{\rm}/g_0$
  outlines the peculiar (``reentrant'') peak shape for intermediate input drive.}
\label{fig:Fig7}
\end{figure}

\section{Conclusion}
In this paper, we have shown how the addition of an additional cross-Kerr
coupling term to the usual radiation-pressure coupling term in the description
of the dynamics of optomechanical systems substantially modifies the stability
properties of the corresponding semiclassical equations of motion. In
particular, we have shown how the inclusion of a CK term leads to an alteration
and, eventually, to the disappearance of the unstable branch
$\mathbf{B}-\mathbf{C}$, for $g_{\rm ck}>g^{\rm c1}_{\rm ck}$. 
Analogously, the unstable branch $\mathbf{D}-\mathbf{E}$ is affected by the
presence of a CK coupling term: for values of $g_{\rm ck} \in \left] 0,  g^*_{\rm
  ck}\right]$ the branch $\mathbf{D}-\mathbf{E}$ is extended ($\mathbf{E} \to \infty$,
for $g_{\rm ck}=g_{\rm ck}^*$ ), while, upon further increase of  $g_{\rm ck}$,
the branch $\mathbf{D}-\mathbf{E}$ shrinks and, eventually, disappears for 
$g_{\rm ck}=g_{\rm ck}^{\rm c2}$

cross-Kerr coupling constant $g_{\rm ck}$ is increased.

\section{Acknowledgements}
\label{sec:acknowledgements}
This work was supported by the Academy of Finland (project ``Quantum properties
of optomechanical systems'').   

\bibliography{./ck_cl}

\begin{thebibliography}{22}%
\makeatletter
\providecommand \@ifxundefined [1]{%
 \@ifx{#1\undefined}
}%
\providecommand \@ifnum [1]{%
 \ifnum #1\expandafter \@firstoftwo
 \else \expandafter \@secondoftwo
 \fi
}%
\providecommand \@ifx [1]{%
 \ifx #1\expandafter \@firstoftwo
 \else \expandafter \@secondoftwo
 \fi
}%
\providecommand \natexlab [1]{#1}%
\providecommand \enquote  [1]{``#1''}%
\providecommand \bibnamefont  [1]{#1}%
\providecommand \bibfnamefont [1]{#1}%
\providecommand \citenamefont [1]{#1}%
\providecommand \href@noop [0]{\@secondoftwo}%
\providecommand \href [0]{\begingroup \@sanitize@url \@href}%
\providecommand \@href[1]{\@@startlink{#1}\@@href}%
\providecommand \@@href[1]{\endgroup#1\@@endlink}%
\providecommand \@sanitize@url [0]{\catcode `\\12\catcode `\$12\catcode
  `\&12\catcode `\#12\catcode `\^12\catcode `\_12\catcode `\%12\relax}%
\providecommand \@@startlink[1]{}%
\providecommand \@@endlink[0]{}%
\providecommand \url  [0]{\begingroup\@sanitize@url \@url }%
\providecommand \@url [1]{\endgroup\@href {#1}{\urlprefix }}%
\providecommand \urlprefix  [0]{URL }%
\providecommand \Eprint [0]{\href }%
\providecommand \doibase [0]{http://dx.doi.org/}%
\providecommand \selectlanguage [0]{\@gobble}%
\providecommand \bibinfo  [0]{\@secondoftwo}%
\providecommand \bibfield  [0]{\@secondoftwo}%
\providecommand \translation [1]{[#1]}%
\providecommand \BibitemOpen [0]{}%
\providecommand \bibitemStop [0]{}%
\providecommand \bibitemNoStop [0]{.\EOS\space}%
\providecommand \EOS [0]{\spacefactor3000\relax}%
\providecommand \BibitemShut  [1]{\csname bibitem#1\endcsname}%
\let\auto@bib@innerbib\@empty
\bibitem [{\citenamefont {Aspelmeyer}\ \emph {et~al.}(2014)\citenamefont
  {Aspelmeyer}, \citenamefont {Kippenberg},\ and\ \citenamefont
  {Marquardt}}]{Aspelmeyer:2014ce}%
  \BibitemOpen
  \bibfield  {author} {\bibinfo {author} {\bibfnamefont {M.}~\bibnamefont
  {Aspelmeyer}}, \bibinfo {author} {\bibfnamefont {T.~J.}\ \bibnamefont
  {Kippenberg}}, \ and\ \bibinfo {author} {\bibfnamefont {F.}~\bibnamefont
  {Marquardt}},\ }\href@noop {} {\bibfield  {journal} {\bibinfo  {journal}
  {Rev. Mod. Phys.}\ }\textbf {\bibinfo {volume} {86}},\ \bibinfo {pages}
  {1391} (\bibinfo {year} {2014})}\BibitemShut {NoStop}%
\bibitem [{Ano(2015{\natexlab{a}})}]{Anonymous:kqVfZnE3}%
  \BibitemOpen
  \href {http://www.ego-gw.it/virgodescription/pag_4.html} {\enquote {\bibinfo
  {title} {{The VIRGO interferometer}},}\ } (\bibinfo {year}
  {2015}{\natexlab{a}})\BibitemShut {NoStop}%
\bibitem [{Ano(2015{\natexlab{b}})}]{Anonymous:j0Ww0aRI}%
  \BibitemOpen
  \href {http://ligo.org/} {\enquote {\bibinfo {title} {{LSC - LIGO Scientific
  Collaboration}},}\ } (\bibinfo {year} {2015}{\natexlab{b}})\BibitemShut
  {NoStop}%
\bibitem [{\citenamefont {Teufel}\ \emph {et~al.}(2011)\citenamefont {Teufel},
  \citenamefont {Donner}, \citenamefont {Li}, \citenamefont {Harlow},
  \citenamefont {Allman}, \citenamefont {Cicak}, \citenamefont {Sirois},
  \citenamefont {Whittaker}, \citenamefont {Lehnert},\ and\ \citenamefont
  {Simmonds}}]{Teufel:2011jga}%
  \BibitemOpen
  \bibfield  {author} {\bibinfo {author} {\bibfnamefont {J.~D.}\ \bibnamefont
  {Teufel}}, \bibinfo {author} {\bibfnamefont {T.}~\bibnamefont {Donner}},
  \bibinfo {author} {\bibfnamefont {D.}~\bibnamefont {Li}}, \bibinfo {author}
  {\bibfnamefont {J.~W.}\ \bibnamefont {Harlow}}, \bibinfo {author}
  {\bibfnamefont {M.~S.}\ \bibnamefont {Allman}}, \bibinfo {author}
  {\bibfnamefont {K.}~\bibnamefont {Cicak}}, \bibinfo {author} {\bibfnamefont
  {A.~J.}\ \bibnamefont {Sirois}}, \bibinfo {author} {\bibfnamefont {J.~D.}\
  \bibnamefont {Whittaker}}, \bibinfo {author} {\bibfnamefont {K.~W.}\
  \bibnamefont {Lehnert}}, \ and\ \bibinfo {author} {\bibfnamefont {R.~W.}\
  \bibnamefont {Simmonds}},\ }\href@noop {} {\bibfield  {journal} {\bibinfo
  {journal} {Nature}\ }\textbf {\bibinfo {volume} {475}},\ \bibinfo {pages}
  {359} (\bibinfo {year} {2011})}\BibitemShut {NoStop}%
\bibitem [{\citenamefont {Andrews}\ \emph {et~al.}(2014)\citenamefont
  {Andrews}, \citenamefont {Peterson}, \citenamefont {Purdy}, \citenamefont
  {Cicak}, \citenamefont {Simmonds}, \citenamefont {Regal},\ and\ \citenamefont
  {Lehnert}}]{Andrews:2014kja}%
  \BibitemOpen
  \bibfield  {author} {\bibinfo {author} {\bibfnamefont {R.~W.}\ \bibnamefont
  {Andrews}}, \bibinfo {author} {\bibfnamefont {R.~W.}\ \bibnamefont
  {Peterson}}, \bibinfo {author} {\bibfnamefont {T.~P.}\ \bibnamefont {Purdy}},
  \bibinfo {author} {\bibfnamefont {K.}~\bibnamefont {Cicak}}, \bibinfo
  {author} {\bibfnamefont {R.~W.}\ \bibnamefont {Simmonds}}, \bibinfo {author}
  {\bibfnamefont {C.~A.}\ \bibnamefont {Regal}}, \ and\ \bibinfo {author}
  {\bibfnamefont {K.~W.}\ \bibnamefont {Lehnert}},\ }\href@noop {} {\bibfield
  {journal} {\bibinfo  {journal} {Nat. Phys.}\ }\textbf {\bibinfo {volume}
  {10}},\ \bibinfo {pages} {321} (\bibinfo {year} {2014})}\BibitemShut
  {NoStop}%
\bibitem [{\citenamefont {Massel}\ \emph {et~al.}(2011)\citenamefont {Massel},
  \citenamefont {Heikkil{\"a}}, \citenamefont {Pirkkalainen}, \citenamefont
  {Cho}, \citenamefont {Saloniemi}, \citenamefont {Hakonen},\ and\
  \citenamefont {Sillanp{\"a}{\"a}}}]{Massel:2011ca}%
  \BibitemOpen
  \bibfield  {author} {\bibinfo {author} {\bibfnamefont {F.}~\bibnamefont
  {Massel}}, \bibinfo {author} {\bibfnamefont {T.~T.}\ \bibnamefont
  {Heikkil{\"a}}}, \bibinfo {author} {\bibfnamefont {J.~M.}\ \bibnamefont
  {Pirkkalainen}}, \bibinfo {author} {\bibfnamefont {S.~U.}\ \bibnamefont
  {Cho}}, \bibinfo {author} {\bibfnamefont {H.}~\bibnamefont {Saloniemi}},
  \bibinfo {author} {\bibfnamefont {P.~J.}\ \bibnamefont {Hakonen}}, \ and\
  \bibinfo {author} {\bibfnamefont {M.~A.}\ \bibnamefont {Sillanp{\"a}{\"a}}},\
  }\href@noop {} {\bibfield  {journal} {\bibinfo  {journal} {Nature}\ }\textbf
  {\bibinfo {volume} {480}},\ \bibinfo {pages} {351} (\bibinfo {year}
  {2011})}\BibitemShut {NoStop}%
\bibitem [{\citenamefont {Rabl}(2011)}]{Rabl:2011gn}%
  \BibitemOpen
  \bibfield  {author} {\bibinfo {author} {\bibfnamefont {P.}~\bibnamefont
  {Rabl}},\ }\href@noop {} {\bibfield  {journal} {\bibinfo  {journal} {Phys.
  Rev. Lett.}\ }\textbf {\bibinfo {volume} {107}},\ \bibinfo {pages} {063601}
  (\bibinfo {year} {2011})}\BibitemShut {NoStop}%
\bibitem [{\citenamefont {Nunnenkamp}\ \emph {et~al.}(2011)\citenamefont
  {Nunnenkamp}, \citenamefont {B{\o}rkje},\ and\ \citenamefont
  {Girvin}}]{Nunnenkamp:2011cp}%
  \BibitemOpen
  \bibfield  {author} {\bibinfo {author} {\bibfnamefont {A.}~\bibnamefont
  {Nunnenkamp}}, \bibinfo {author} {\bibfnamefont {K.}~\bibnamefont
  {B{\o}rkje}}, \ and\ \bibinfo {author} {\bibfnamefont {S.}~\bibnamefont
  {Girvin}},\ }\href@noop {} {\bibfield  {journal} {\bibinfo  {journal} {Phys.
  Rev. Lett.}\ }\textbf {\bibinfo {volume} {107}},\ \bibinfo {pages} {063602}
  (\bibinfo {year} {2011})}\BibitemShut {NoStop}%
\bibitem [{\citenamefont {Heikkil{\"a}}\ \emph {et~al.}(2014)\citenamefont
  {Heikkil{\"a}}, \citenamefont {Massel}, \citenamefont {Tuorila},
  \citenamefont {Khan},\ and\ \citenamefont
  {Sillanp{\"a}{\"a}}}]{Heikkila:2014hh}%
  \BibitemOpen
  \bibfield  {author} {\bibinfo {author} {\bibfnamefont {T.~T.}\ \bibnamefont
  {Heikkil{\"a}}}, \bibinfo {author} {\bibfnamefont {F.}~\bibnamefont
  {Massel}}, \bibinfo {author} {\bibfnamefont {J.}~\bibnamefont {Tuorila}},
  \bibinfo {author} {\bibfnamefont {R.}~\bibnamefont {Khan}}, \ and\ \bibinfo
  {author} {\bibfnamefont {M.~A.}\ \bibnamefont {Sillanp{\"a}{\"a}}},\
  }\href@noop {} {\bibfield  {journal} {\bibinfo  {journal} {Phys. Rev. Lett.}\
  }\textbf {\bibinfo {volume} {112}},\ \bibinfo {pages} {203603} (\bibinfo
  {year} {2014})}\BibitemShut {NoStop}%
\bibitem [{\citenamefont {Pirkkalainen}\ \emph {et~al.}(2015)\citenamefont
  {Pirkkalainen}, \citenamefont {Cho}, \citenamefont {Massel}, \citenamefont
  {Tuorila}, \citenamefont {Heikkil{\"a}}, \citenamefont {Hakonen},\ and\
  \citenamefont {Sillanp{\"a}{\"a}}}]{Pirkkalainen:2015hh}%
  \BibitemOpen
  \bibfield  {author} {\bibinfo {author} {\bibfnamefont {J.~M.}\ \bibnamefont
  {Pirkkalainen}}, \bibinfo {author} {\bibfnamefont {S.~U.}\ \bibnamefont
  {Cho}}, \bibinfo {author} {\bibfnamefont {F.}~\bibnamefont {Massel}},
  \bibinfo {author} {\bibfnamefont {J.}~\bibnamefont {Tuorila}}, \bibinfo
  {author} {\bibfnamefont {T.~T.}\ \bibnamefont {Heikkil{\"a}}}, \bibinfo
  {author} {\bibfnamefont {P.~J.}\ \bibnamefont {Hakonen}}, \ and\ \bibinfo
  {author} {\bibfnamefont {M.~A.}\ \bibnamefont {Sillanp{\"a}{\"a}}},\
  }\href@noop {} {\bibfield  {journal} {\bibinfo  {journal} {Nat Commun}\
  }\textbf {\bibinfo {volume} {6}},\ \bibinfo {pages} {6981} (\bibinfo {year}
  {2015})}\BibitemShut {NoStop}%
\bibitem [{\citenamefont {Seok}\ \emph {et~al.}(2013)\citenamefont {Seok},
  \citenamefont {Buchmann}, \citenamefont {Wright},\ and\ \citenamefont
  {Meystre}}]{Seok:2013iw}%
  \BibitemOpen
  \bibfield  {author} {\bibinfo {author} {\bibfnamefont {H.}~\bibnamefont
  {Seok}}, \bibinfo {author} {\bibfnamefont {L.~F.}\ \bibnamefont {Buchmann}},
  \bibinfo {author} {\bibfnamefont {E.~M.}\ \bibnamefont {Wright}}, \ and\
  \bibinfo {author} {\bibfnamefont {P.}~\bibnamefont {Meystre}},\ }\href@noop
  {} {\bibfield  {journal} {\bibinfo  {journal} {Phys. Rev. A}\ }\textbf
  {\bibinfo {volume} {88}},\ \bibinfo {pages} {063850} (\bibinfo {year}
  {2013})}\BibitemShut {NoStop}%
\bibitem [{\citenamefont {Jayich}\ \emph {et~al.}(2008)\citenamefont {Jayich},
  \citenamefont {Sankey}, \citenamefont {Zwickl}, \citenamefont {Yang},
  \citenamefont {Thompson}, \citenamefont {Girvin}, \citenamefont {Clerk},
  \citenamefont {Marquardt},\ and\ \citenamefont {Harris}}]{Jayich:2008iz}%
  \BibitemOpen
  \bibfield  {author} {\bibinfo {author} {\bibfnamefont {A.~M.}\ \bibnamefont
  {Jayich}}, \bibinfo {author} {\bibfnamefont {J.~C.}\ \bibnamefont {Sankey}},
  \bibinfo {author} {\bibfnamefont {B.~M.}\ \bibnamefont {Zwickl}}, \bibinfo
  {author} {\bibfnamefont {C.}~\bibnamefont {Yang}}, \bibinfo {author}
  {\bibfnamefont {J.~D.}\ \bibnamefont {Thompson}}, \bibinfo {author}
  {\bibfnamefont {S.~M.}\ \bibnamefont {Girvin}}, \bibinfo {author}
  {\bibfnamefont {A.~A.}\ \bibnamefont {Clerk}}, \bibinfo {author}
  {\bibfnamefont {F.}~\bibnamefont {Marquardt}}, \ and\ \bibinfo {author}
  {\bibfnamefont {J.~G.~E.}\ \bibnamefont {Harris}},\ }\href@noop {} {\bibfield
   {journal} {\bibinfo  {journal} {New J Phys}\ }\textbf {\bibinfo {volume}
  {10}},\ \bibinfo {pages} {095008} (\bibinfo {year} {2008})}\BibitemShut
  {NoStop}%
\bibitem [{\citenamefont {Purdy}\ \emph {et~al.}(2010)\citenamefont {Purdy},
  \citenamefont {Brooks}, \citenamefont {Botter}, \citenamefont {Brahms},
  \citenamefont {Ma},\ and\ \citenamefont {Stamper-Kurn}}]{Purdy:2010gh}%
  \BibitemOpen
  \bibfield  {author} {\bibinfo {author} {\bibfnamefont {T.}~\bibnamefont
  {Purdy}}, \bibinfo {author} {\bibfnamefont {D.}~\bibnamefont {Brooks}},
  \bibinfo {author} {\bibfnamefont {T.}~\bibnamefont {Botter}}, \bibinfo
  {author} {\bibfnamefont {N.}~\bibnamefont {Brahms}}, \bibinfo {author}
  {\bibfnamefont {Z.~Y.}\ \bibnamefont {Ma}}, \ and\ \bibinfo {author}
  {\bibfnamefont {D.}~\bibnamefont {Stamper-Kurn}},\ }\href@noop {} {\bibfield
  {journal} {\bibinfo  {journal} {Phys. Rev. Lett.}\ }\textbf {\bibinfo
  {volume} {105}},\ \bibinfo {pages} {133602} (\bibinfo {year}
  {2010})}\BibitemShut {NoStop}%
\bibitem [{\citenamefont {Xuereb}\ and\ \citenamefont
  {Paternostro}(2013)}]{Xuereb:2013ug}%
  \BibitemOpen
  \bibfield  {author} {\bibinfo {author} {\bibfnamefont {A.}~\bibnamefont
  {Xuereb}}\ and\ \bibinfo {author} {\bibfnamefont {M.}~\bibnamefont
  {Paternostro}},\ }\href@noop {} {\bibfield  {journal} {\bibinfo  {journal}
  {PRA}\ }\textbf {\bibinfo {volume} {87}},\ \bibinfo {pages} {023830}
  (\bibinfo {year} {2013})}\BibitemShut {NoStop}%
\bibitem [{\citenamefont {Nemoto}\ and\ \citenamefont
  {Munro}(2004)}]{Nemoto:2004dr}%
  \BibitemOpen
  \bibfield  {author} {\bibinfo {author} {\bibfnamefont {K.}~\bibnamefont
  {Nemoto}}\ and\ \bibinfo {author} {\bibfnamefont {W.~J.}\ \bibnamefont
  {Munro}},\ }\href@noop {} {\bibfield  {journal} {\bibinfo  {journal} {Phys.
  Rev. Lett.}\ }\textbf {\bibinfo {volume} {93}},\ \bibinfo {pages} {250502}
  (\bibinfo {year} {2004})}\BibitemShut {NoStop}%
\bibitem [{\citenamefont {Sheng}\ \emph {et~al.}(2012)\citenamefont {Sheng},
  \citenamefont {Zhou}, \citenamefont {Zhao},\ and\ \citenamefont
  {Zheng}}]{Sheng:2012jd}%
  \BibitemOpen
  \bibfield  {author} {\bibinfo {author} {\bibfnamefont {Y.-B.}\ \bibnamefont
  {Sheng}}, \bibinfo {author} {\bibfnamefont {L.}~\bibnamefont {Zhou}},
  \bibinfo {author} {\bibfnamefont {S.-M.}\ \bibnamefont {Zhao}}, \ and\
  \bibinfo {author} {\bibfnamefont {B.-Y.}\ \bibnamefont {Zheng}},\ }\href@noop
  {} {\bibfield  {journal} {\bibinfo  {journal} {Phys. Rev. A}\ }\textbf
  {\bibinfo {volume} {85}},\ \bibinfo {pages} {012307} (\bibinfo {year}
  {2012})}\BibitemShut {NoStop}%
\bibitem [{\citenamefont {Sheng}\ and\ \citenamefont
  {Zhou}(2015)}]{Sheng:2015dz}%
  \BibitemOpen
  \bibfield  {author} {\bibinfo {author} {\bibfnamefont {Y.-B.}\ \bibnamefont
  {Sheng}}\ and\ \bibinfo {author} {\bibfnamefont {L.}~\bibnamefont {Zhou}},\
  }\href@noop {} {\bibfield  {journal} {\bibinfo  {journal} {Sci. Rep.}\
  }\textbf {\bibinfo {volume} {5}},\ \bibinfo {pages} {7815} (\bibinfo {year}
  {2015})}\BibitemShut {NoStop}%
\bibitem [{\citenamefont {Sheng}\ \emph {et~al.}(2010)\citenamefont {Sheng},
  \citenamefont {Deng},\ and\ \citenamefont {Long}}]{Sheng:2010ky}%
  \BibitemOpen
  \bibfield  {author} {\bibinfo {author} {\bibfnamefont {Y.-B.}\ \bibnamefont
  {Sheng}}, \bibinfo {author} {\bibfnamefont {F.-G.}\ \bibnamefont {Deng}}, \
  and\ \bibinfo {author} {\bibfnamefont {G.~L.}\ \bibnamefont {Long}},\
  }\href@noop {} {\bibfield  {journal} {\bibinfo  {journal} {Phys. Rev. A}\
  }\textbf {\bibinfo {volume} {82}},\ \bibinfo {pages} {032318} (\bibinfo
  {year} {2010})}\BibitemShut {NoStop}%
\bibitem [{\citenamefont {Walls}\ and\ \citenamefont
  {Milburn}(2007)}]{Walls:1105914}%
  \BibitemOpen
  \bibfield  {author} {\bibinfo {author} {\bibfnamefont {D.~F.}\ \bibnamefont
  {Walls}}\ and\ \bibinfo {author} {\bibfnamefont {G.~J.}\ \bibnamefont
  {Milburn}},\ }\href@noop {} {\emph {\bibinfo {title} {{Quantum Optics}}}}\
  (\bibinfo  {publisher} {Springer},\ \bibinfo {address} {Berlin},\ \bibinfo
  {year} {2007})\BibitemShut {NoStop}%
\bibitem [{\citenamefont {Aldana}\ \emph {et~al.}(2013)\citenamefont {Aldana},
  \citenamefont {Bruder},\ and\ \citenamefont {Nunnenkamp}}]{Aldana:2013iya}%
  \BibitemOpen
  \bibfield  {author} {\bibinfo {author} {\bibfnamefont {S.}~\bibnamefont
  {Aldana}}, \bibinfo {author} {\bibfnamefont {C.}~\bibnamefont {Bruder}}, \
  and\ \bibinfo {author} {\bibfnamefont {A.}~\bibnamefont {Nunnenkamp}},\
  }\href@noop {} {\bibfield  {journal} {\bibinfo  {journal} {Phys. Rev. A}\
  }\textbf {\bibinfo {volume} {88}},\ \bibinfo {pages} {043826} (\bibinfo
  {year} {2013})}\BibitemShut {NoStop}%
\bibitem [{\citenamefont {Marquardt}\ \emph {et~al.}(2007)\citenamefont
  {Marquardt}, \citenamefont {Chen}, \citenamefont {Clerk},\ and\ \citenamefont
  {Girvin}}]{Marquardt:2007dn}%
  \BibitemOpen
  \bibfield  {author} {\bibinfo {author} {\bibfnamefont {F.}~\bibnamefont
  {Marquardt}}, \bibinfo {author} {\bibfnamefont {J.~P.}\ \bibnamefont {Chen}},
  \bibinfo {author} {\bibfnamefont {A.~A.}\ \bibnamefont {Clerk}}, \ and\
  \bibinfo {author} {\bibfnamefont {S.~M.}\ \bibnamefont {Girvin}},\
  }\href@noop {} {\bibfield  {journal} {\bibinfo  {journal} {Phys. Rev. Lett.}\
  }\textbf {\bibinfo {volume} {99}},\ \bibinfo {pages} {093902} (\bibinfo
  {year} {2007})}\BibitemShut {NoStop}%
\bibitem [{\citenamefont {Khan}\ \emph {et~al.}(2015)\citenamefont {Khan},
  \citenamefont {Massel},\ and\ \citenamefont {Heikkil{\"a}}}]{Khan:2015kf}%
  \BibitemOpen
  \bibfield  {author} {\bibinfo {author} {\bibfnamefont {R.}~\bibnamefont
  {Khan}}, \bibinfo {author} {\bibfnamefont {F.}~\bibnamefont {Massel}}, \ and\
  \bibinfo {author} {\bibfnamefont {T.~T.}\ \bibnamefont {Heikkil{\"a}}},\
  }\href@noop {} {\bibfield  {journal} {\bibinfo  {journal} {Phys. Rev. A}\
  }\textbf {\bibinfo {volume} {91}},\ \bibinfo {pages} {043822} (\bibinfo
  {year} {2015})}\BibitemShut {NoStop}%
\end{thebibliography}%

\end{document}